# Racism in the Machine: Visualization Ethics in Digital Humanities Projects

Katherine Hepworth <khepworth_at_unr_dot_edu>, University of Nevada, Reno
Christopher Church <christopherchurch_at_unr_dot_edu>, University of Nevada, Reno

## Abstract

Data visualizations are inherently rhetorical, and therefore bias-laden visual artifacts that contain both explicit and implicit arguments. The implicit arguments depicted in data visualizations are the net result of many seemingly minor decisions about data and design from inception of a research project through to final publication of the visualization. Data workflow, selected visualization formats, and individual design decisions made within those formats all frame and direct the possible range of interpretation, and the potential for harm of any data visualization. Considering this, it is imperative that we take an ethical approach to the creation and use of data visualizations. Therefore, we have suggested an ethical data visualization workflow with the dual aim of minimizing harm to the subjects of our study and the audiences viewing our visualization, while also maximizing the explanatory capacity and effectiveness of the visualization itself. To explain this ethical data visualization workflow, we examine two recent digital mapping projects, Racial Terror Lynchings and Map of White Supremacy Mob Violence.

## Introduction

[1] In March 2016, Microsoft's Technology and Research division released an artificial intelligence chatbot known as TayTweets under the handle @TayandYou on Twitter. Microsoft had programmed the chatbot to interact with the online community and subsequently learn from the tweets of users mentioning it, in order to improve its natural language communication abilities. Though unlikely to pass the Turing Test any time soon, TayTweets was nevertheless able to generate its own internet memes, crack its own jokes, and participate in the regular back-and-forth of online conversation.

[2] Microsoft presented TayTweets as a value-neutral project that would showcase the technological achievements of its research division, and early on the AI was seemingly innocuous, posting anodyne responses to tweets welcoming it to Twitter and asking it mundane questions. With its first tweet of "hellooooooo w¿¿¿¿rld!!!" — a riff on the traditional output of a computer programmer's first program — TayTweets had virtually stepped onto the world stage and greeted the Twitterverse in a seemingly innocent, even youthfully naive, way.

[3] Within 16 hours, however, the experiment had to be shut down, as TayTweets had algorithmically learned to be racist. Posting inflammatory tweets championing white supremacy and denigrating racial minorities and marginalized groups, the artificial intelligence program had become an online menace and in fact a cyberbully. TayTweets's rapid descent into racist demagoguery serves as a harrowing reminder that our digital productions are not free from the cultural assumptions and prejudices that shape everyday human experience. Microsoft's designers had a blind spot about the depth and breadth of American racism, which allowed TayTweets to replicate deep-seated preconceptions and prejudices without critically examining them.

[4] For our purposes, what is interesting about TayTweets is not its experimentation with artificial intelligence, but the project's apparent assumption that an algorithm — the set of rules by which any programmatic approach operates, whether digital or analog — is value neutral and divorceable from human prejudice and malice. On the contrary, humans

are at the center of algorithms, not only as their creators but, in the case of data-driven algorithms, as the producers of the content they shape and present. Though an extreme example, TayTweets clearly demonstrates how wider cultural assumptions, prevalent political ideologies, and public discourses shape the output of our algorithmic productions, potentially replicating what we already know instead of aiding us to discover the new and uncover the unexamined. In other words, TayTweets tapped into an American zeitgeist riddled with detrimental preconceptions with regard to race.

Ethical visualization is essentially a human-centric approach to algorithmic production, considering the underlying biases, ideologies, and beliefs that animate algorithms as they structure and reproduce past inequities and harmful realities. All data visualizations — whether static or interactive, printed or digital, computer-generated or hand-drawn — are algorithmic by nature, in the sense that they solve complex problems of representation and require a set of tree-based actions and decisions for their success. Ethical visualization practices sit at the intersection of humanistic inquiry, ethics, and communication design. We define ethical visualization as the presentation of visualized information in ways that acknowledge and mitigate the potential for harm engendered within the visualization form and content. While good design practice forms the backbone of ethical visualizations, ethical visualization practice goes one step further to consider the ultimate societal impact of such design choices: do such choices cause harm or mislead, either intentionally or unintentionally? Do they result in a net societal benefit, or do they prove deleterious to marginalized individuals? These questions must be brought to the forefront when considering good design, because a visualization can follow good design practices and consequently be easy to understand, but still produce a negative societal impact for its subject matter all the same.

Considering the object lesson of TayTweets, this article proposes that digital humanists must adhere to a form of visualization ethics that considers how both choices about working with data and the rhetorical qualities of communication elements — color, composition, line, symbols, type, and interactivity — shape users' understandings of represented people and places. Given the "racism in the machine" — the ways in which our digital tools can inadvertently recreate the latent racism, underlying prejudices, and cultural blind spots in our society — the goal of this visualization ethics should be "increasing understanding [for users] while minimizing harm" to represented people and places [Cairo 2014].

To propose a methodology for visualization ethics, this article examines two projects that visualize the horrific history of racial lynching in the United States, Lynching in America by the Equal Justice Initiative and Monroe Work Today by Auut Studio, in order to show that no visualization is ideologically neutral, but is instead part of an argument that must be critically examined. In contrast to TayTweets's outright bigotry, Lynching in America and Monroe Work Today both present visualizations that demonstrate to varying degrees ethical visualization practices, though one succeeds in producing an ethical visualization to a greater degree than the other. This paper showcases ethical visualization in practice to varying degrees in these two digital mapping projects, demonstrating how choices about representation, interaction, and annotation in their data visualizations either do harm in the sense described above, or challenge dominant narratives. In comparing these two projects, the article outlines a workflow that can ensure that data visualizations adhere to best practices in visualization ethics and thereby present opportunities for more inclusive and critical interaction with represented data.

## Lynching in America (EJI) vs Monroe Work Today

### Example 1: Racial Terror Lynchings Map

Lynching in America (https://lynchinginamerica.eji.org/), a promotional website made by Google for the Equal Justice Initiative, a mass incarceration not-for-profit organization, contains an interactive map titled Racial Terror Lynchings (https://lynchinginamerica.eji.org/explore). Drawing on a dataset compiled by the EJI for their eponymous report of "4075 racial terror lynchings of African-Americans," this interactive choropleth map of the United States purports to depict "Reported lynchings by county", occurring between 1877 and 1950 ["Lynching in America"]. Overall, the map has a minimalist aesthetic, reminiscent of Google's Material Design visual style (https://material.io/) that depicts topographical elevation as well as country, state, and county borders, but not cities, towns, roads, rivers, lakes, or landmarks (see Figure 1). The data visualized is also minimalist in presentation, and is tied directly to the state and county borders depicted, as is typical of choropleth maps.

The map is focused on the southeastern United States. On first page load in screen widths below 1500px, the map centers on the United States below the Mason-Dixon line, while in larger screens the entire contiguous United States is shown. However, in both cases, the American south grabs attention with many southern counties highlighted. Users can click on any county or state, and are taken to a zoomed-in view of the state, with the total reported lynchings in that state displayed in large letters. The user can then hover over individual counties to find out how many lynchings were reported in that county (see Figure 2). This zoomed-in view focuses on the number of lynchings per state and county — represented by polygons shaped according to county boundaries — creating a visual argument that encourages measurement of states and counties as more or less reprehensible in terms of number of lynchings. It is focused on lynchings in the context of political boundaries, and thereby presents a strong geopolitical argument about recorded lynchings that appears definitive and damning, particularly of the states with the most counties marked in bright red: Alabama, Florida, Louisiana, and Mississippi. There is a straightforwardness about this visual argument: racial terror was and is morally wrong, and its contours are plain to any competent observer.

In terms of color, the map features a dark, almost monochrome color palette, with a dark grey United States segmented by black state lines, placed in a dark blue sea. This mostly dark color scheme is dramatically interrupted with many counties highlighted in various shades of red, with bright red indicating 20 or more lynchings recorded in the county (see Figure 1). The color scheme of red on an otherwise dark, monochrome palette compounds the visual argument described above, as it references the brutality and violence inflicted upon African Americans in those locations, recalling the blood stains on United States history as it relates to racial violence and white supremacy.

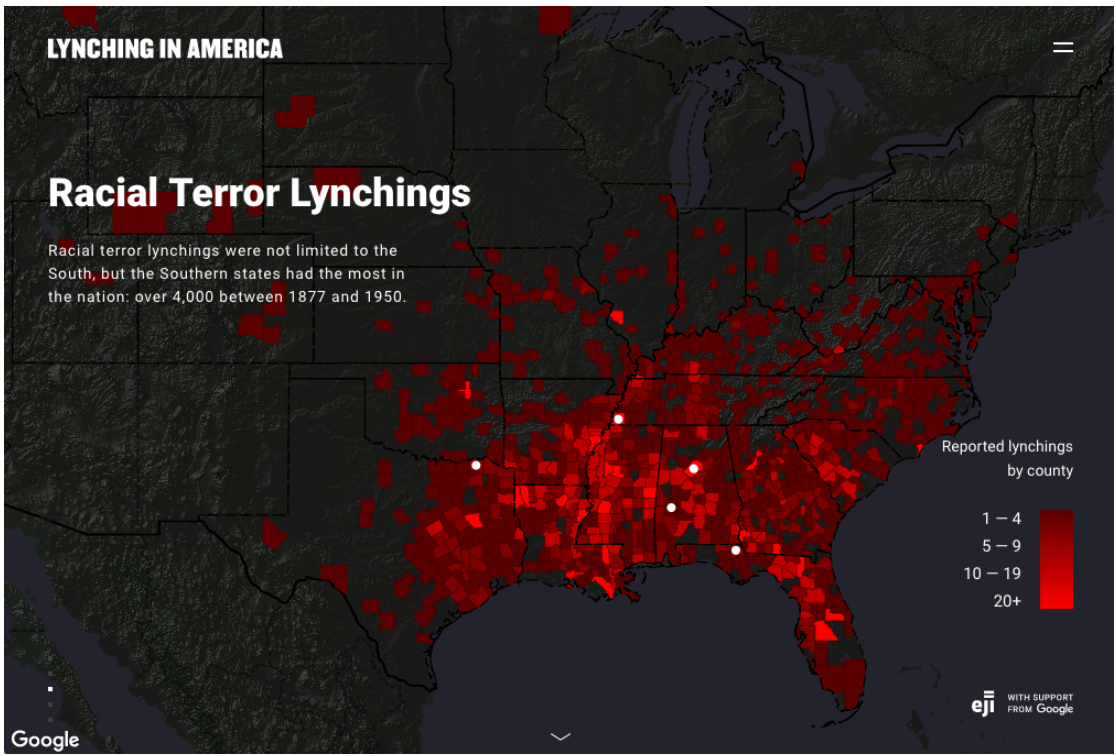

**Figure 1.** Figure 1

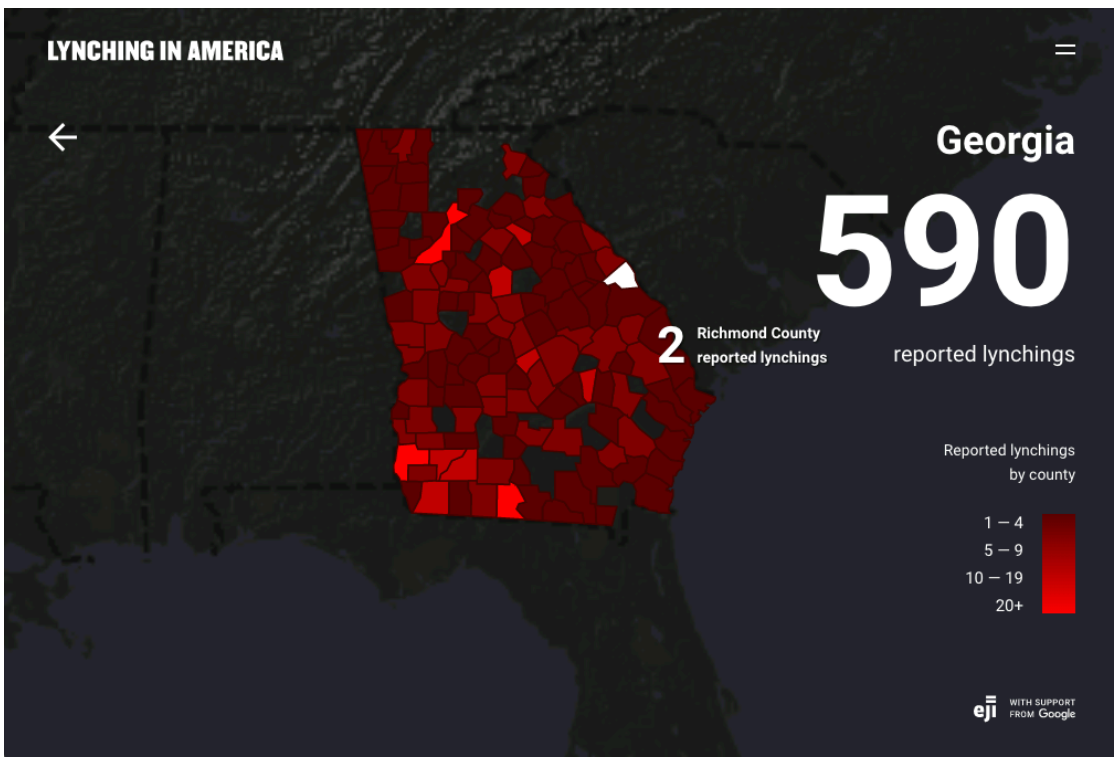

**Figure 2.** Racial Terror Lynchings map in Lynching in America website. Top image (Figure 1) shows initial view of map on page load. Bottom image (Figure 2) shows the zoomed-in view on Georgia, with mouse hover over Richmond County. With 590 reported lynchings in Georgia, Richmond County's 2 reported lynchings appear significantly less reprehensible than the record of other counties in Georgia.

## Example 2: Map of White Supremacy Mob Violence

The second example, Map of White Supremacy Mob Violence (http://www.monroeworktoday.org/explore/), is a far more complex visualization than the minimalist Racial Terror Lynchings. It is an interactive map within Monroe Work Today (http://www.monroeworktoday.org), a website dedicated to publicizing the research of sociologist Monroe Work, who systematically documented lynchings in the United States. Created by education-focused digital agency Auut Studio, this interactive map of the United States depicts lynching records in the context of historical racial violence and public discourses of white superiority, and is consequently subtitled "The lynchings and riots to enforce racial superiority in the US". Upon first page load, Map of White Supremacy Mob Violence stands in stark contrast to Racial Terror Lynchings, presenting the entire contiguous United States center of screen, irrespective of screen size, and representing each recorded case of lynching as a single grey dot on a white national landmass background (see Figure 3). State borders are not visible in this view, and a single, bright, attention grabbing color is only found on instructions for users. This initial map view — including map framing, choice of colors, and the carefully worded title and subtitle — provides a strong and markedly different message than the Racial Terror Lynchings map. Collectively, these elements present a strong visual argument that discourses of white supremacy are a nationwide reality in the United States, one that has historically been enforced through lynchings and riots by mobs. The geopolitical framing of lynching as a nation-wide reality is balanced with the acknowledgement of a crucial aspect of the historically pervasive intellectual climate, one of white superiority, and an important aspect of the social behavioral climate, mob violence. At the center bottom of the page is a key annotation: a label that reads "Should I trust this? Find out." If users click this message, they are taken to a plethora of information about the veracity of the data used, and a discussion of the importance of thinking critically about the visualized data.

Although it is not entirely clear from this initial page view what the grey dots represent, the user is presented with instructions to zoom in and click on individual points. After the user follows these instructions, the map gains more color, and individual dots become more visible, along with state and county boundaries (see Figure 4). In this view, state and county lines are visible, as are the boundaries of Native American reservations and areas that historically were Spanish colonies. However, these geographic lines are indicated with subtlety, using pale colors. In contrast, the dots representing each individual lynching are highlighted using bright colors and an *onclick* interaction effect. Marking geopolitical boundaries but deemphasizing them, combined with emphasizing each individual lynching with bright colors, makes the visual argument that lynchings occurred in the context of geopolitical boundaries, but that the individual deaths are of greater significance than the boundaries themselves.

Along the bottom of the screen, supplemental information and interactive features appear, including a timeline, a label stating the time span represented in the current view, and a color-coded legend of dot colors. Six different dot colors are used: five to represent races of lynched people, and one to represent "other", where the race recorded in the historical records does not fall into one of the main five. Notably, the "other" category includes lynchings of white abolitionists, thereby demonstrating the complexity of the history of lynching in the United States, and that white people were also, however rarely, victims of lynching in the name of white supremacy.

Both the timeline and the legend serve two purposes: one functional, the other persuasive. Users can select a time period using the timeline, which then alters data presented on the map, so they can see how many lynchings occurred in an area over a specific timeframe. Secondly, the timeline provides a persuasive visual cue that racial superiority-motivated lynchings occured continually over a long timespan, with some time periods seeing evidence of significantly more racial superiority-motivated lynchings. The legend also works on these two levels. It firstly allows the user to identify the race of a lynched person based on the color of the dot used to represent them, and secondly provides a strong visual counterargument to the widespread public assumption that lynchings were perpetrated exclusively on African Americans. These two elements, the timeline and the legend, confront the user with the temporal and racial extent of white superiority-motivated lynchings, both qualities that are absent from the Racial Terror Lynchings map.

Most strikingly, the zoomed-in view of Map of White Supremacy Mob Violence contains a list, in the bottom left hand corner of the screen, of the name (where available) of every lynched person represented by a dot in the current map view, and the year they were lynched. Clicking on any individual dot brings up a *callout* box containing extra details of the lynching available in the historical record, including the county in which it occurred, the details of mob violence in which the lynching occurred, the accusation made before the lynching, and links to every available historical source that verifies the record (see Figure 5). Naming individuals who were lynched, and providing circumstances surrounding their death, focuses attention on the humanity of victims of lynching and on the social circumstances in which lynching was a viable possibility. In the present day context of many white supremacists denying their racism, it is worth noting that these historical records rarely mention race as a motivator for lynching. For example, William B. Willis of Richmond County, Georgia, was accused of murder before being lynched. The Map of White Supremacy Mob Violence does an exemplary job of demonstrating that racism was indeed the motivating factor in lynchings, and also that racism was largely hidden in the official historical record by documenting other, non-racial reasons. Providing links to multiple historical sources within the interactive map increases trust in the veracity of data, while at the same time giving users the opportunity to investigate the historical evidence themselves. Map of White Supremacy Mob Violence uses multiple compelling strategies to both humanize the data it represents, and to contextualize it in the societal racism and discourses of white supremacy in the United States.

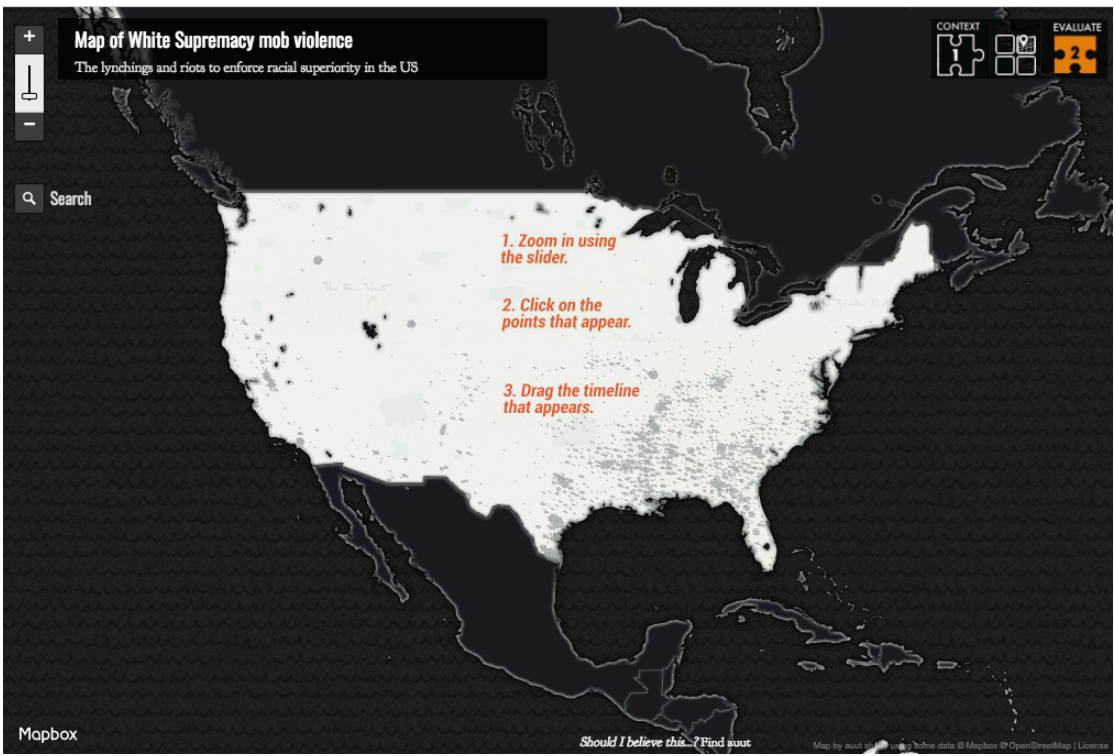

**Figure 3.** Figure 3

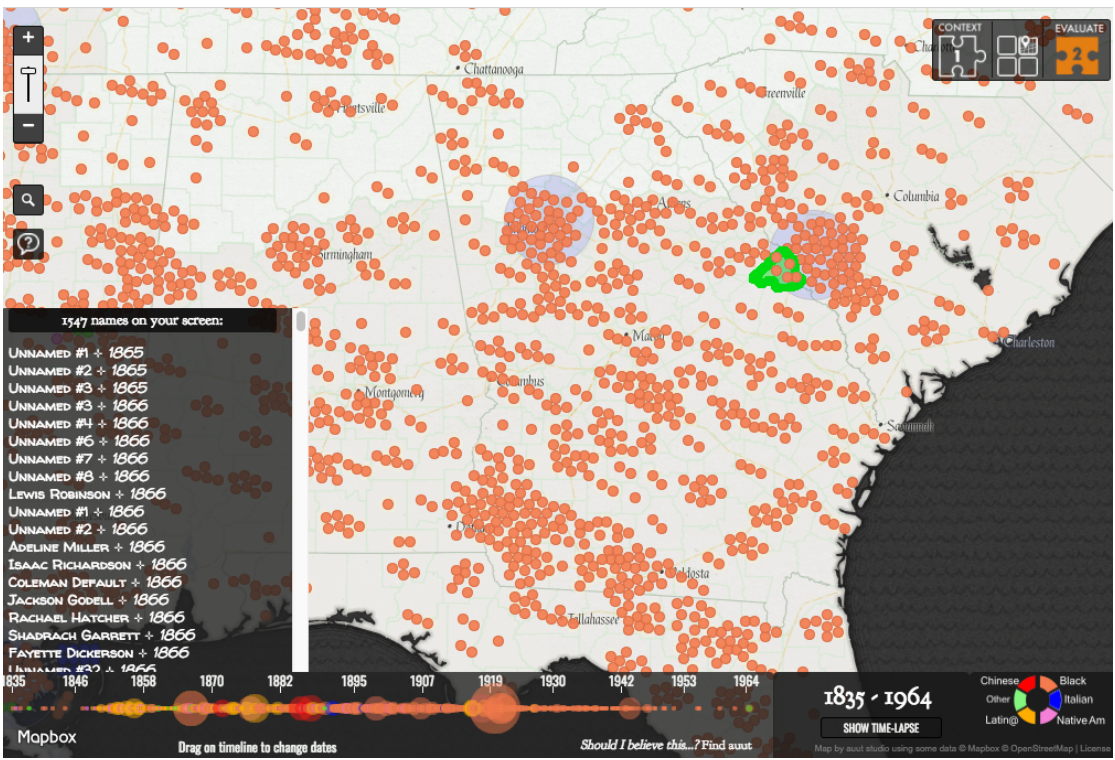

**Figure 4.** Figure 4

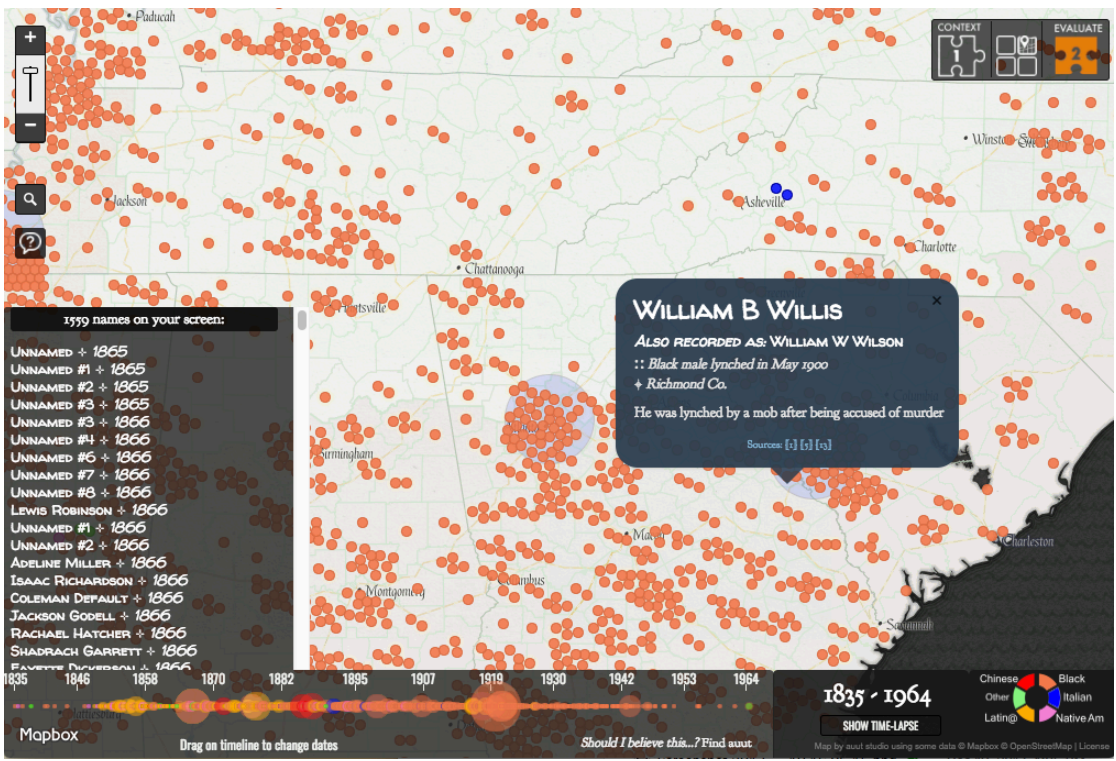

**Figure 5.** Map of White Supremacy Mob Violence in Monroe Work Today website. Top image (Figure 3) shows the initial view of the map on page load. Center image (Figure 4) shows a zoomed-in view on Georgia. Richmond County is highlighted in green by the authors, for comparison with the Racial Terror Lynching map in Figures 1 and 2. Bottom image (Figure 5) shows a view of clicking over an individual lynching record in Richmond County. Naming individuals who were lynched and providing circumstances of their death focuses attention on the humanity of victims of lynching, the cover up of racist motivations for lynching documented in the historical record, and deemphasizes geopolitical comparisons. Additionally, the context of a large number of lynchings close across the border in South Carolina suggests that the geographic area including Richmond county was no less immune to racial superiority-motivated violence than the surrounding areas, as the presentation of information in the Racial Terror Lynchings map visually argues.

## Comparison of both maps' depictions of the West

The contrast between these two maps is even more striking when looking at the west coast of the United States. For example, Figure 6 shows a marked difference between the recorded lynchings in California in the Racial Terror Lynchings map and the Map of White Supremacy Mob Violence respectively. The difference in the representation can be accounted for by the fact that the former only depicts lynchings of African Americans, whereas the latter depicts lynchings of African Americans, Native Americans, Latinos, Italians, and other races. The view of the American West depicted in Map of White Supremacy Mob Violence in Figure 6b provides a compelling visual narrative that lynchings were common across California in the name of enforcing white superiority. The view of Racial Terror Lynchings Map depicted in Figure 6a, by its lack of clarity about exactly which data is being represented (i.e. historical records of lynchings of African Americans only), its use of a bold and expansive title that suggests comprehensive coverage of lynchings of all races (i.e. Racial Terror Lynchings), and its tonal emphasis on the south (California appearing grey, while visual attention is drawn to the large amounts of red in the bottom right hand corner of the map), makes a visual argument that California had few instances of racial terror or lynchings. This is particularly problematic because the Equal Justice Initiative's stated goal is to challenge black incarceration, and a key part of their organizational message is that this goal is urgent and directly related to the history of lynchings motivated by white supremacy. Black incarceration rates in California are among the highest in the nation today: according to the U.S. Bureau of Justice, California imprisons blacks 8.8 times more frequently than whites, well above the national average of 5.5:1 ["The Sentencing Project"]. Consequently EJI's visual argument in the Racial Terror Lynchings Maps inadvertently breaks down in advocacy regarding California prisons.

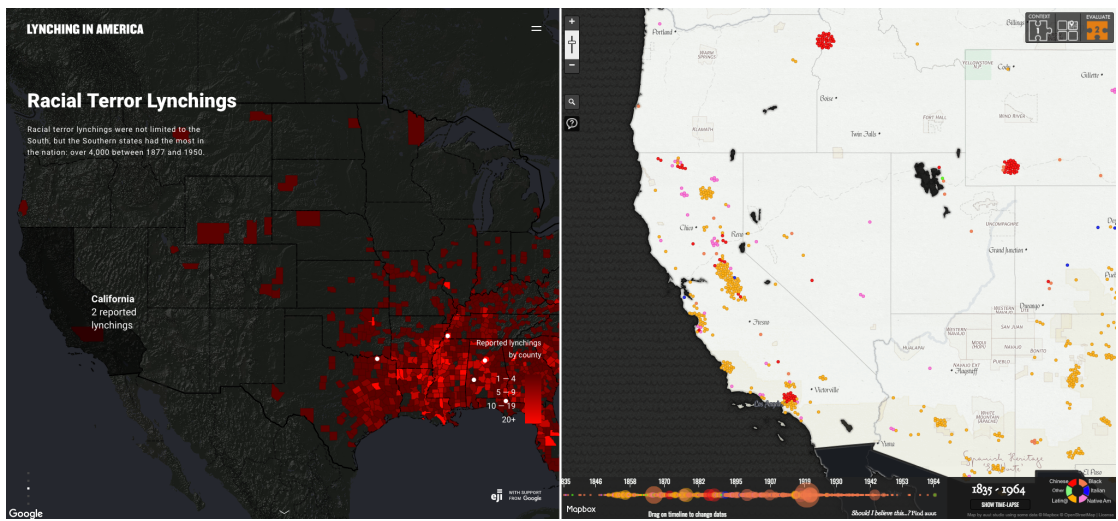

**Figure 6.** Contrasting views of California in Racial Terror Lynchings Map (figure 6a, left) and Map of White Supremacy Mob Violence (figure 6b, right). The map framing on the left suggests lynching occurred rarely in California, whereas the map framing on the right indicates that lynching was widespread in California.

## Discussion

The emphases of these two maps are necessarily different because of the different purposes of the sites in which they are situated. Lynching in America is a promotional and advocacy tool for the Equal Justice Initiative, primarily created to visualize data within (and thereby promote) the report "Lynching in America", which records lynchings of African Americans and frames lynching as a societal tool — enabled through mob violence and discourses of white superiority — to subjugate African Americans between slavery and mass incarceration. It is in the Equal Justice Initiative's interests to visualize historical lynching data in a way that draws attention to geopolitical divides, so that clear links can be made between historical lynching events and present-day constituencies of sitting politicians as well as county and state local governments. The map provides compelling visual evidence for the organization's present-day advocacy work regarding the inequitable mass incarceration of black Americans, the case of California notwithstanding. The problematic aspect of this is that the Equal Justice Initiative's website, report, and Racial Terror Lynchings map are unfortunately named to suggest that they cover all historical records of lynchings in the United States. The Lynching in America report includes instructions for educators who wish to use it as a teaching resource. In this context, the geopolitical emphasis, use of color, use of summary data, and lack of links to sources give a concerning impression that African-American lynchings were the complete record of lynchings in the United States for the purposes of racial terror.

The majority of both scholarship and public attention regarding lynchings centers on the experience of African Americans in the Southern United States, and for good measure: for blacks in the Jim Crow South, lynchings represented a terrifying aspect of everyday life. In the grand scheme of racial violence, however, lynchings represented one small piece in a complex puzzle of individual, institutional, and structural racism. The conflation of lynching with the full extent of racialized violence in United States history obscures the historic depth and breadth of the oppression of people of color. A black individual was far more likely to suffer public humiliation, assault, rape, and murder than a public lynching. While lynchings do not represent the totality of racial violence in America, they come to the fore because they were highly symbolic affairs: gruesome spectacles of white supremacy, racial violence, and bodily mutilation meant to suppress and intimidate as much as they were meant to kill [Wood 2009, 1–4]. Due to their highly symbolic nature and the lasting implications of racist attitudes, policies, and actions for African Americans today, lynchings have become synonymous with racial hatred in the postbellum American south.

18

However, mob murder historically extended well beyond Dixie, representing a form of prejudicial frontier "justice" in the Midwest, West, and Southwest against minorities and immigrants of various backgrounds [Pfeifer 2006] [Pfeifer 2013]. Outside of the latest scholarship, such victims, whom Carrigan and Webb describe as the "forgotten dead," are largely overshadowed or overlooked in the public sphere, missing an opportunity to explore the structural, cross-regional, and

19

transethnic dimensions of American lynching. In fact, lynch mobs murdered hundreds of Mexicans between 1848 and 1928 in the American Southwest [Carrigan and Webb 2017].

Moving beyond EJI's limited focus on African-American populations in the American South, Auut Studio acknowledges the historic violence committed against Native American populations, which is noticeably absent from the vast majority of lynching data sets. This general oversight plays into present-day blind spots regarding violence against native populations, who, despite suffering more state violence and community disruption than any other minority group in 2016 according to data collected by the Centers for Disease Control and Prevention, rarely garner the public spotlight ["The Counted"]. To counteract the public oversight of this "forgotten minority," Auut Studio included the boundaries of Native American reservations as "sovereignt[ies] deserving of equal visual treatment on the map" [Ramey 2017]. Similarly, Auut Studio included lynchings of the Chinese along the frontier, namely in California, as well as Mexicans in the Southwest, keeping in line with recent historical scholarship.

In order to ethically represent historical subjects, data visualization techniques, particularly those geared toward public consumption, must remain abreast of the insights made in two areas of the scholarly literature: debates on the subjects they depict, and debates on design and representational considerations in the ethical visualization literature. Present-day academic debates on lynching challenge the widely accepted notion that lynching was exclusively a Southern phenomenon, excusing those regions of the United States outside of the South of their own racist heritage. The small but growing ethical visualization literature emphasizes the need for acknowledging and mitigating the potential for harm inherent in visualizing data, particularly when it comes to selection of design elements, visual style, and selections of data to annotate and visually emphasize [Cairo 2014] [Hepworth 2016] [Kostelnick 2016] [Skau, Harrison, and Kosara 2015].

Collectively, we tend to visualize old arguments, as visualization practices have not kept pace with dominant arguments in the digital humanities literature about the importance of critical practices. Visualization practice in the digital humanities runs the risk of following a functionalist methodological approach that assumes visualization to be an impartial medium. This illusory functionalism has led others to charge that data collection, processing, and visualization practices constitute mere "janitorial work" in the service of "real" humanities scholarship, ignoring the important decisions made during such processes that critically shape historical narratives. Construing digital humanities practice as a "support field" has led to further accusations that the digital humanities simply show to us what we already know, rather than challenging us to think critically about historical topics in new, interesting, and socially responsible ways [Allington et al. 2016].

Following Alan Liu's charge that digital humanists have ignored cultural criticism, which in turn has blocked "the digital humanities from becoming a full partner of the humanities," data visualization practitioners need to critically engage with the ways in which digital tools can "communicate humanity" rather than relegating it to the margins, or worse, obscuring the human stories essential to understanding structural racism today [Liu 2012].

Visualizing data that exclusively focuses on the African-American experience in the Southern United States provides an important argument about the nature of Jim Crow racism. However, purporting such data to be an inclusive representation constitutes a harm in the sense that it perpetuates common narratives of racial violence as a southern exception to an otherwise inclusive nation. After publishing their interactive map, the Equal Justice Initiative itself recognized this oversight, acknowledging the 300 lynchings of African Americans outside of the American south, though notably leaving aside other ethnicities like Native Americans, Mexicans, and the Chinese that fell victim to much of Western and Southwestern mob violence ["EJI Releases New Data" 2017]. This overlooks the depth of structural racism and its support of white supremacy, thereby denying the experience of millions of present-day Americans. Historical information has the capacity to legitimize or delegitimize present-day experience, and visualizations of historical data are a particularly compelling and resonant medium through which such information can either harm or help.

## Ethical Visualization Workflow

We call for critical and practical analysis of the entire endeavor of data collection and visualization in the digital humanities. Humanities scholars in recent decades have critically examined the categories of scientific analysis inherited from the enlightenment that presuppose essential differences (based on sex, race, age etc) acknowledging that such pre-suppositions frame and ultimately determine scholarly insight [Knorr-Cetina 1981]. However, digital humanists and data scientists rely heavily on these categories in their visualization practices precisely because they animate the entirety of the scientific endeavor.

Similarly, visual communication has been studied for decades in terms of its highly rhetorical qualities [Barton and Barton 1985] [Gallagher et al. 2011] [Tapia and Hodgkinson 2003]. Despite early interventions by journalist Darrell Huff and statistician Howard Wainer, data visualization literature and practice seldom focus on the argument-altering, persuasive qualities of individual design decisions or visualization conventions to a degree that allows for effectively mitigating the harmful potential consequences of visualization [Huff 1954] [Wainer 1984]. One notable exception to this overall trend is the work of cartography scholar Mark Monmonier, who has long advocated for acknowledging the complexity and nuance inherent in the minutiae of visualization design decisions [Monmonier 1991] [Monmonier 1995]. Huff, Monmonier, and Wainer can be seen as the grandfathers of a small, interdisciplinary body of work on ethical visualization practices that directly tackles the challenge of mitigating the potential for harm inherent in data visualization [Cairo 2014] [Hepworth 2016] [Kostelnick 2007]. We argue that there is an urgent need for this ethical visualization literature to grow in detail and scope, particularly with regard to digital humanities projects. It is imperative that ethical data visualizers evaluate not only the rhetorical decisions of the analysis, but also critically examine the entire process of working with data from collection to final visualization and publishing.

Visual theorist Johanna Drucker offers one proposal to address the challenge of ethical representation. She argues that the humanities need their own forms of visualization, distinct from those developed for administrative and scientific purposes [Drucker 2011, 1]. She does this for good reason: the standard visualization conventions that we are most familiar with — bar charts, line charts, pie charts — were all created in European countries at the height of their colonial expansion and industrial transformation. They were created to track demographics, trade, war, and debt; all the trappings of their growing empires [Wainer 2013] [Cole 2000]. These visualization conventions carry this history, and these associations, with them.

However, in his work on the role of charts in the social sciences, historian Howard S. Becker reminds us that "if we invent a new format every time we have something to say, we risk alienating users" [Becker 2007, 169]. Finding the right balance between visualization innovation and working within established conventions is a complex procedure that demands a combination of high visual literacy, advanced visualization production skills, intimate understanding of the visualization context, and a critical perspective on the entire data collection and visualization process. Much valuable work has been done by geographers in terms of working critically with established visualization formats, in the form of critical GIS [Harvey et al. 2005] [Thatcher 2016].

We argue that for pragmatic reasons, humanists must work with the visualization formats that are familiar to their audiences much of the time. We encourage innovation in visualization practices only insofar as innovations are both intelligible to the intended audience, and that they foster consideration of the dignity of the represented subjects. Therefore, we propose an ethical visualization workflow (see Figure 7) that operates within existing data collection and information design frameworks but ensures that any given visualization's argument provides a compelling yet ethical and accurate representation of historical subjects.

Prior to creating a data visualization, a scholar following our ethical visualization workflow would complete several critical steps: defining, reviewing, collecting, pruning, describing, surveying, and pre-visualizing. These steps involve processes that many digital humanities scholars will be familiar with, with the important difference that they are suggested here with alterations that we believe will result in an ethical data visualization. The steps can be grouped into three standard digital humanities practice phases: pre-data collection (defining, reviewing); data collection and curation (collecting, pruning, describing); and data visualization and argumentation (surveying, pre-visualizing, visualizing, publishing).

## Pre-Data Collection

In the pre-data collection phase, the first step involves clearly defining the subject area that the data visualization will depict, while the second step involves reviewing the latest secondary literature on the topic at hand. Reviewing subject area literature would inform the remaining steps in the ethical data visualization workflow, inviting the researcher to compensate for the data set's shortcomings by seeking out and including new information, or to limit the scope of the visual argument to be produced with said data. Doing so would avoid the glaring oversights and interpretive overreach that plagued the EJI's Racial Terror Lynchings Map.

## Data Collection and Curation

The second phase, data collection and curation, is perhaps most crucial in producing an ethical data visualization, precisely because it is so frequently overlooked. The third step in the ethical visualization workflow involves collecting primary documents, artifacts, and datasets, as well as secondary datasets of potential relevance, while the fourth step involves checking the appropriateness, authenticity, veracity, and feasibility of use of collected primary and secondary materials, and pruning those that don't hold up under scrutiny. Once these two critical data collection steps have been finished, the researcher completes the fifth step, describing, by creating their own dataset that combines the collected materials. [32]

Creating a custom dataset for the researcher's visualization in this way eliminates other people's and institutions' biases from the data, ensuring erroneous arguments are not unintentionally added through using unaltered historical datasets. This process of collecting, pruning, and describing data sets was undertaken by Auut Studios for ten years before visualization, contributing to the particularly considerate treatment of ethical factors in the Map of White Supremacy Mob Violence. Similarly, EJI created an extensive dataset of over 4,000 public lynchings based on work done by Tuskegee University and the research of E.M. Beck and Stewart E. Tolnay that compellingly shows the long legacy of terroristic violence in the American south. Nevertheless, decisions made about what counts or does not count as "racial terror violence" made during the data collection phase — namely to exclude the American frontier — ultimately shaped EJI's visual argument, making it overreach in its claims and thereby creating a narrative around racial violence that excludes other minorities and other geographic locales. [33]

| PHASES | STEPS |
|---|---|
| Pre-data collection | 1. Defining field of inquiry |
| | 2. Reviewing latest subject scholarship |
| Data curation and collection | 3. Collecting primary documents and artifacts |
| | 4. Pruning non-viable primary documents |
| | 5. Describing primary data in custom data set |
| Data visualizing and argumentation | 6. Surveying ethical visualization literature |
| | 7. Pre-visualization context consideraton |
| | 8. Visualizing data |
| | 9. Publishing visualization |

**Figure 7.** The Ethical Visualization Workflow we propose for producing visualizations that minimize harm to three groups: people using visualizations, people represented in visualizations, and people personally affected by the represented material.

## Data Visualization and Argumentation

The third and final phase of the visualization workflow, data visualization and argumentation, is the main one associated with the ethics of data visualization, and this phase involves four steps: surveying, pre-visualizing, visualizing, and publishing.

**Surveying & Pre-visualizing**

Once the custom data set has been created, the researcher moves onto the sixth step: surveying the latest literature from the small but growing interdisciplinary field of ethical visualization. Surveying this literature allows the researcher to keep abreast of ethical visualization innovations and recommended best practices. The seventh step, pre-visualizing, involves considering the contextual factors around the visualization: normalizing representations of the data; selecting the publishing medium; identifying intended and potential unintended audiences based on that medium; and discerning between visualization formats possible in that medium. When making decisions about the argument produced by the visualization, it is imperative to consider the larger contextual framing and the story that it tells. For instance, in the case of EJI's Racial Terror Lynchings, the data isn't normalized against census population data. Normalizing the data in this way would show how frequent these lynchings were in a population of people rather than a geographic space, thus showing the relative societal impact of a lynching of two people in a county of 100 rather than 5 people in a county of 10,000.

In considering these contextual factors, the researcher can then select the most appropriate visualization format, creating test visualizations (these are more rudimentary than, and distinct from, alpha prototypes) and performing any necessary re-structuring of the dataset based on findings of this prototyping. In the pre-visualizing step, the selection of visualization format is particularly important. The mechanics and conventions of specific visualization formats contribute to determining meaning. For example, Mercator geographic projections have been criticized for privileging Northern Europe and underemphasizing the Global South [Monmonier 2010]. While they receive less attention for their distortional effects, pie charts encourage a comparison between visualized elements as if they make up a unified whole, whether or not they actually do so in reality [Tufte 1998]. It is not that such visualization formats are by definition unethical to use, but that a critical perspective can unearth the ways in which they limit or close off possibilities of argumentation.

[36]

Critically examining each of the areas involved in pre-visualizing, and making careful, considered choices on each area, can dramatically change the ethical implications of a project. For example, if the Lynching in America report and Racial Terror Map were printed documents, they would have a much smaller audience than they have as web-based documents. This would result in less reach for the organizations, but also less potential harm in terms of the Racial Terror Lynching map being used in contexts outside of lobbying for prison reform, and therefore giving an erroneous presentation of the history of lynching in the United States.

[37]

The pre-visualizing step provides an opportunity to acknowledge the prior understanding and cultural frame of the intended and unintended potential audiences. The persuasive and culturally bound associations those audiences necessarily have with design elements, explanatory text, headers, legends and interaction experiences need to be considered. The choice of colors and color ramps, as well as graphic or cartographic elements like political boundaries, invariably influence the argument produced by the visualization, as do map default views at certain screen widths, and zoom options. To be ethical, these choices must be made with the scholarly literature and the ethical visualization literature in mind, as well as a critical perspective on the power of individual design elements. To maintain visualization ethics, they should strive to minimize harm while increasing understanding, and this can only be done when the latest ethical developments in the field are factored into the visualization at hand.

[38]

For instance, given the research on the dehumanization of deaths by aggregating them into faceless statistics and the resulting inability to enact meaningful social change [Du Wors et al. 1960] [Bernard et al. 1971] [Slovic 2007] [Katz 2011], Monroe Work Today created an individual marker for every person killed, so that their name and details could be uncovered by the viewer. While more visually complex than a choropleth map, as was used in Racial Terror Lynchings, creating individual markers conveys the sheer gravity of the violence while not losing sight of the individuals who suffered at the hands of white supremacy. As the director of Auut Studio, RJ Ramey, explains,

[39]

> it was [a] conscious decision to include every person killed as their own marker on the map — so that their name could be discovered. I have received suggestions that a more "efficient" visualization would have been a choropleth map or graduated symbol size — and visualizers at EJI and the NYT have utilized these methods. However, for purposes of respecting the gravity of the violence and the humanity of the victims, I very much believed it was most appropriate to provide the audience a census of the lynching record, not a visual or numerical analysis. [Ramey 2017]

In contrast, EJI replicates the conventions, and thereby the limitations, of how the Tuskegee Institute has visualized their data by state and county since 1931 (see, for example: *Lynchings by States and Counties*, 1931, https://www.loc.gov/resource/g3701e.ct002012).

[40]

The visual argument produced must be a bounded one, explaining what it does show while inviting the user to interrogate and explore what it does not show. Effacing the data, along with its assumptions, within the visualization itself, as Map of White Supremacy Mob Violence so admirably does, invites the user to judge the veracity and scope of the data themselves, providing an opportunity for users to build informed trust in the visualized data. Auut Studio makes plain the decisions made during all of these pre-visualization steps, encouraging the user to take a critical stance toward the argument ultimately put forth by the visualization itself. On the other hand, while EJI does provide a link to its full lynching report, which goes into detail about how the dataset was collected and why it was bounded in such a way as to exclude the American frontier, this link stands apart from the visualization itself, buried within the website's navigation drawer. Moreover, the report does not address the translation of those choices into the rhetorical qualities of the visualization itself, which purports to show the entirety of American Racial Terror Lynchings. As the argument presenting the data, an ethical visualization should provide clear and apparent options for users to investigate sources, so that the user will be able to get a critical sense of the underlying data set. Showing the data effectively equates to showing your work for the user, but it does not necessitate providing the user with what amount to false choices that obfuscate the rhetorical value of the visualization in the first place.

[41]

**Visualizing and Publishing**

The eighth step involves development of the visualization itself. To be as ethical as possible, this needs to be an iterative process, beginning with an alpha prototype, ending with a final visualization, and including rounds of user testing with intended and unintended audiences after each round of iteration. While user testing is not commonplace in the digital humanities, it is a long-established practice in the allied fields of computer science and visual communication design [Nielsen 1993] [Sanders and Stappers 2012]. Such rigorous prototyping and testing ensures the research mitigates harm to audiences that may result from drawing associations and conclusions that are false, and that the researcher could not predict. Lastly, the ninth step involves publishing the visualization, the culmination of a thorough and considered ethical research practice. Finally, in the interest of reproducibility and data interrogation, it is vital to publish alongside the visualization its underlying datasets or, in the case of lacking the requisite rights, to provide ample documentation and citation of those datasets.

[42]

**Feasibility of Proposed Workflow**

Digital humanities teams can implement this workflow by centering their activity on ethical questions around their subject area and the technology used to present it. This can be accomplished using a twofold approach: firstly, by familiarizing themselves with the latest research in the content field and adjacent fields; secondly, by including team members familiar with the entirety of the data pipeline from collection to cleaning to presentation, as well as communication design principles and user experience methods. User experience design is particularly important for evaluating the interpretive intervention made by the visualization and mitigating harm caused by the final visualization.

[43]

We recognize that the combination of skills we advocate is not the norm in digital humanities projects and that it is rare for any one single humanist to possess all of these skills. However, these are skills that are common in several disciplines, particularly in the humanities (source interrogation and intellectual framing); social sciences (data collection, curation, and analysis); and communication design departments (interface design and user experience). Interdisciplinary collaboration in teams that contain — or consult with — humanists, social scientists, and communication designers is the most feasible way to implement the ethical visualization workflow. There is strong interest in digital humanities collaborations in the field of communication design, as evidenced by a recent special issue of *Visible Language* (see volume 49, issue 3), and there are exemplary collaborations between social scientists and humanists that can serve as models for such teams (see the cooperation between the Digital Humanities and the Social Science D-Lab at University of California, Berkeley).

[44]

Ethical data visualization is much more about priorities and planning in digital projects than about increasing the amount of resources, or using the latest technology. Whereas EJI's Lynching in America had the full support of Google Labs, Auut Studio's Monroe Work Today resulted from the careful consideration of a single individual working in consultation with domain-area experts. Auut Studio began as a single-person operation, but through proper planning, interdisciplinary collaboration, and attention to historiographical implications, it was able to present a more ethically minded visualization of lynching data sets.

[45]

Digital humanists need to bring their skepticism toward their source material to bear on visualizations themselves, and consultation with communication design faculty will illuminate the design and interaction elements that need to be interrogated. The medium matters as much as the content in shaping the message, and thus it is essential to understand the medium in order to fully appreciate a visualization's societal intervention. In other words, as we stated at the outset of this paper, the technology is not itself value-neutral.

[46]

# Conclusion

Data visualizations are inherently rhetorical, and therefore bias-laden visual artifacts that contain both explicit and implicit arguments. The implicit arguments depicted in data visualizations are the net result of many seemingly minor decisions about data and design from inception of a research project through to final publication of the visualization. Data workflow, selected visualization formats, and individual design decisions made within those formats all frame and direct the possible range of interpretation, and the potential for harm of any data visualization.

47

Considering this, it is imperative that we take an ethical approach to the creation and use of data visualizations. Therefore, we have suggested an ethical data visualization workflow — defining, reviewing, collecting, pruning, describing, surveying, pre-visualizing, visualizing, and publishing — with the dual aim of minimizing harm to the subjects of our study and the audiences viewing our visualization, while also maximizing the explanatory capacity and effectiveness of the visualization itself. To arrive at our ethical data visualization workflow, we have examined two recent digital mapping projects, Racial Terror Lynchings and Map of White Supremacy Mob Violence, to demonstrate the potential pitfalls of data visualization, as well as suggest ethical ways to avoid such pitfalls.

48

While EJI's Racial Terror Lynchings is an admirable project, it nevertheless presents an incomplete picture of racial lynchings in the American South in a way that forecloses meaningful discussions about racism and white supremacy in the American North and West, leaving out the "forgotten dead" among frontier populations. By contrast, Monroe Work Today's Map of White Supremacy Mob Violence is an example of the possibilities of ethical visualization, precisely because of the extended period in which the data were interrogated in line with the latest scholarship in the field. Auut Studio brought a critical eye to the topic at hand, acknowledging the shortcomings of the data and investing the time to ultimately create a more inclusive picture of white supremacist violence.

49

Monroe Work Today's Map of White Supremacy Mob Violence is an exemplar of ethical visualization, not because it is free of two centuries of baggage and biases, but because it acknowledges them, while also acknowledging the potential pitfalls of the very endeavor of transforming human beings into visualized historical data. Unlike TayTweets, who consumed the racist prejudices and ideologies of our modern society and became an intolerable digital bigot, Monroe Work Today gives real thought to the quality and veracity of the data, the ways in which that data represented (and did not fully stand in for) marginalized persons, and the design decisions taken in visually representing that data. In so doing, it presents an accessible, nuanced and compelling account of America's sordid history of lynching those on the margins. Likewise, by following our process, readers can create similarly critical and self-effacing visualizations that make apparent the argument, assumptions, and inherent flaws that animate their digital humanities projects.

50

## Notes

[1] Both authors contributed equally to this article.